\documentclass[12pt,english]{extarticle}
\pdfoutput=1
\usepackage[charter]{mathdesign}

\usepackage[T1]{fontenc}
\usepackage[latin9]{inputenc}
\usepackage[a4paper]{geometry}
\usepackage{color}
\usepackage{babel}
\usepackage{float}
\usepackage{booktabs}
\usepackage{url}
\usepackage{amsmath}
\usepackage{microtype}
\usepackage[unicode=true,pdfusetitle,
 bookmarks=true,bookmarksnumbered=false,bookmarksopen=true,bookmarksopenlevel=2,
 breaklinks=false,pdfborder={0 0 0},pdfborderstyle={},backref=false,colorlinks=true]
 {hyperref}
\hypersetup{
 urlcolor=purple,citecolor=purple,linkcolor=purple}

\makeatletter

\newcommand{\noun}[1]{\textsc{#1}}
\providecommand{\tabularnewline}{\\}

\newcommand{\lyxrightaddress}[1]{
	\par {\raggedleft \begin{tabular}{l}\ignorespaces
	#1
	\end{tabular}
	\vspace{1.4em}
	\par}
}
\providecommand*{\code}[1]{\texttt{#1}}

\@ifundefined{date}{}{\date{}}
\usepackage[dvipsnames]{xcolor}
\usepackage{listings}
\usepackage{textcomp}
\input{all.tikzdefs}
\tikzstyle{blank}=[fill=white, shape=circle, draw=white, inner sep=0.8pt]
\tikzstyle{dot}=[fill=black, shape=circle, draw=black, inner sep=0.8pt]

\tikzstyle{edge}=[-, draw=EdgeColor, densely dashed,line width=1.2pt, line cap=rect, style=whitebg]
\tikzstyle{massive1 edge}=[-, draw=Massive1EdgeColor, line width=2.0pt, style=whitebg, line cap=rect]
\tikzstyle{massive2 edge}=[-, draw=Massive2EdgeColor, line width=2.0pt, style=whitebg, line cap=rect]
\tikzstyle{massive3 edge}=[-, draw=Massive3EdgeColor, line width=2.0pt, style=whitebg, line cap=rect]

\tikzstyle{dot1}=[-, draw=none, postaction=decorate, decoration={markings,mark=at position .50 with {\node[style=dot]{};}}]
\tikzstyle{dot2}=[-, draw=none, postaction=decorate, decoration={markings,mark=between positions 0.33 and 0.67 step 0.33 with {\node[style=dot]{};}}]
\tikzstyle{dot3}=[-, draw=none, postaction=decorate, decoration={markings,mark=between positions 0.25 and 0.76 step 0.25 with {\node[style=dot]{};}}]
\tikzstyle{dot4}=[-, draw=none, postaction=decorate, decoration={markings,mark=between positions 0.20 and 0.81 step 0.20 with {\node[style=dot]{};}}]

\usepackage{prettyref}
\let\origref\ref
\newrefformat{chap}{\hyperref[#1]{Chapter~\origref*{#1}}}
\newrefformat{app}{\hyperref[#1]{Appendix~\origref*{#1}}}
\newrefformat{sec}{\hyperref[#1]{Section~\origref*{#1}}}
\newrefformat{subsec}{\hyperref[#1]{Section~\origref*{#1}}}
\newrefformat{tab}{\hyperref[#1]{Table~\origref*{#1}}}
\newrefformat{fig}{\hyperref[#1]{Figure~\origref*{#1}}}
\newrefformat{eq}{eq.\,\hyperref[#1]{(\origref*{#1})}}
\AtBeginDocument{\renewcommand*{\ref}[1]{\prettyref{#1}}}

\newcommand*{\manFl}[1]{\texttt{#1}}
\newcommand*{\manCm}[1]{\texttt{\textbf{#1}}}
\newcommand*{\manAr}[1]{\textit{#1}}
\newcommand*{\manQl}[1]{``\texttt{#1}''}
\newcommand*{\manMa}[1]{$#1$}

\usepackage[absolute]{textpos}
\AtBeginDocument{\begin{textblock*}{3cm}[1,0](18cm,2cm)\raggedleft\texttt{\small KA-TP-26-2022 P3H-22-109}\end{textblock*}}

\makeatother

\usepackage{listings}
\usepackage[bibstyle=numeric,citestyle=numeric-comp,sorting=none]{biblatex}

\begin{document}
\title{\textsc{\textbf{Rational Tracer}}:\\
a Tool for Faster Rational Function Reconstruction}
\author{Vitaly Magerya}
\maketitle

\lyxrightaddress{{\small{}Institute for Theoretical Physics (ITP),}\\
{\small{}Karlsruhe Institute of Technology~(KIT),}\\
{\small{}76128~Karlsruhe, Germany}\\
{\small{}E-mail: \href{mailto:vitalii.maheria@kit.edu}{vitalii.maheria@kit.edu}}\\
{\small{}\hspace{8cm}}}
\begin{abstract}
Rational Tracer (\noun{Ratracer}) is a tool to simplify complicated
arithmetic expressions using modular arithmetics and rational function
reconstruction, with the main idea of separating the construction
of expressions (via \emph{tracing}, i.e. recording the list of operations)
and their subsequent evaluation during rational reconstruction.

\noun{Ratracer} can simplify arithmetic expressions (provided as
text files), solutions of linear equation systems (specifically targeting
Integration-by-Parts (IBP) relations between Feynman integrals), and
even more generally: arbitrary sequences of rational operations, defined
in C++ using the provided library \code{ratracer.h}. Any of these
can also be automatically expanded into series prior to reconstruction.

This paper describes the usage of \noun{Ratracer} specifically focusing
on IBP reduction, and demonstrates its performance benefits by comparing
with \noun{Kira}+\noun{FireFly} and \noun{Fire6}.
\end{abstract}
\tableofcontents{}

\section{Introduction}

Reconstruction of rational functions based on modular arithmetic methods
is an established field of mathematics. Its main idea is that a rational
function in multiple variables can be reconstructed from the knowledge
of its values modulo a prime number with the variables set to fixed
numbers. Because modular evaluation is a comparatively fast operation,
these methods enable one to sidestep the complexity and the intermediate
expression swell associated with simplifying the expressions symbolically
(i.e. using the approaches of the traditional computer algebra systems).

As an illustration of a simpler case (polynomial reconstruction),
with the knowledge of
\begin{equation}
f(11)=606\left(\mathrm{mod}\,997\right)\qquad\text{and}\qquad f(38)=827\left(\mathrm{mod}\,997\right)
\end{equation}
one can use Lagrange interpolation and rational number reconstruction~\cite{Wang:1982,Monagan:2004}
to conclude that 
\begin{equation}
f(x)=606\frac{x-38}{11-38}+827\frac{x-11}{38-11}=996+599x=-1+\frac{4}{5}x\left(\mathrm{mod}\,997\right).
\end{equation}
This form can then be double-checked (or possibly improved) with more
evaluations of $f(x)\,\mathrm{mod}\,n$. Similar methods have also
been developed for rational functions~\cite{Kaltofen:1990,Grigoriev:1990,Grigoriev:1991,deKleine:2005,Khodadad:2006,Kaltofen:2007,Cuyt:2011,Huang:2017},
with the goal of minimizing the number of ``probes'' (black-box
evaluations of the target function modulo a prime).

Rational function reconstruction is actively used in high-energy physics
following the work of~\cite{vonManteuffel:2014ixa,Peraro:2016wsq}
as a method of solving large systems of linear equations generated
as a part of solving Integration-by-Parts (IBP) relations between
the Feynman integrals using the Laporta algorithm~\cite{Laporta:2000dsw}.
Multiple state-of-the-art IBP solvers are using modular arithmetics:
\noun{Fire6}~\cite{Smirnov:2019qkx}, \noun{Kira}~\cite{Maierhofer:2017gsa,Klappert:2020nbg}
when used with \noun{FireFly}~\cite{Klappert:2019emp,Klappert:2020aqs},
\noun{FiniteFlow}~\cite{Peraro:2019svx}, \noun{Caravel}~\cite{Abreu:2020xvt},
and \noun{Finred} (a private implementation by von~Manteuffel et~al).
Solving IBP relations is one of the bottlenecks for computing higher
order perturbative corrections in high-energy physics, and a major
motivation for \noun{Ratracer}.

\section{The \protect\noun{Ratracer} approach}

A typical IBP-solving tool based on modular arithmetics generates
the needed linear equations, and then repeatedly solves them via Gaussian
elimination modulo a 63-bit prime with the variables set to large
random values. The (integer) values of the coefficients in the final
row-reduced form are used to reconstruct them as rational expression.

It is our observation that a typical IBP solver spends the majority
of its time---as much as 90\% of it---managing its data structures
used to represent the equations, rather than performing modular arithmetics.
Fortunately, it is possible to forgo these data structures altogether:
if instead of solving the linear system each time anew, one would
trace it and record each arithmetic operation performed during the
Gaussian elimination (as e.g. \emph{``add location~1 to location~2 and store into location~3''}),
then for subsequent evaluations there is no need to re-run the original
algorithm---only this list of operations. We shall call such a list
``\emph{a trace}''. Evaluating a trace instead of the original
algorithm immediately eliminates the overhead of managing data structures,
and allows one to store all the needed intermediate values in a tightly
packed linear region of memory, improving the performance further.

\noun{Ratracer} is an implementation of this idea. It consists of
a small C++ library, \code{ratracer.h}, used to record traces of
arbitrary user-defined calculations, and the command-line tool, \code{ratracer},
used to simplify and reconstruct traces, as well as parse arithmetic
expressions and solve linear systems of equations.

The trace files \noun{Ratracer} works on are stored in a custom binary
format optimized for performance and convenience; essence the format
simply contains a list of arithmetic operations with the output values
specially marked. For example, an expression like ``$2\times x+3$''
could be encoded as

\begin{lstlisting}
location #1 = integer 2
location #2 = variable 1
location #3 = product of location #1 and #2
location #4 = integer 3
location #5 = sum of location #3 and #4
save location #5 as output 1
\end{lstlisting}

Having recordings of operations as a standalone object comes with
multiple advantages. With a trace file \noun{Ratracer} can:
\begin{itemize}
\item Optimize a trace by eliminating operations that don't contribute to
the output (\emph{``dead code''}, e.g. redundant equations), folding
constants (i.e. replacing $2\times3$ by~$6$), simplifying operations
(i.e. replacing $1\times x$ by just~$x$), merging common subexpressions,
and reshuffling the temporary locations to minimize the memory usage.
\item Reconstruct a trace with a subset of its variables set to integers,
rationals, or arbitrary expressions.
\item Select a subset of a trace's outputs and drop the rest, so that instead
of reconstructing the whole set of resulting coefficients it would
be possible to reconstruct any subset. The idea here is that evaluating
a subset is faster than evaluating the whole set (after dead code
elimination); this also allows to easily parallelize the whole reconstruction
across multiple computers: each can work on reconstructing its own
subset of outputs. This way one can achieve an equivalent of master-wise
reduction, sector-wise reduction, or any mixture of these by just
choosing which outputs to keep in which trace files.
\item Expand the outputs of a trace in a series (without performing reconstruction).
Because the outputs in a trace file are just trees of arithmetic expressions,
any operation that makes sense on arithmetic expressions makes sense
on a trace file too. Expanding a trace in a series has two immediate
uses:
\begin{itemize}
\item automatically calculating derivatives of IBP coefficients without
reconstructing them;
\item increasing the reconstruction performance by expanding IBP coefficients
into series in a small parameter (e.g. the dimensional regulator~$\varepsilon$,
or a small mass ratio), and throwing away higher orders---if those
are deemed to not be needed in practice.
\end{itemize}
\end{itemize}
The main disadvantage is the size of the trace files: because it grows
with the number of the required operations, for problems of interest
it can go into the gigabyte range, making it impractical to keep it
in the memory. For this reason \noun{Ratracer} makes sure that:
\begin{enumerate}
\item During trace recording (i.e. expression tracing or equations solving)
the operations are not kept in the memory, but are progressively written
to disk instead.
\item During optimization transformations the whole trace is never fully
loaded into the memory---it is read in chunks from the disk instead---and
the memory requirement of temporary data structures used during the
transformation is never proportional to the number of operations,
but is at most proportional to the memory size required to evaluate
the trace (which is always assumed to fit into the memory).

This makes the algorithms used for these optimizations more complicated,
and limits transformations like common expression elimination to work
on a subset of the code at a time, rather than being global. Still,
\noun{Ratracer} manages to work within this constraint well.
\item During the evaluation the trace is never fully loaded into memory
(unless specifically allowed by the user), and is instead read in
chunks from the disk.

We have found that the time to read a very large trace from a solid-state
drive is less than the time needed to evaluate the corresponding trace,
so the overhead is not great, but the parallel evaluation is limited
by this for larger traces. We do not however believe this to be a
principal limitation, as additional ways to improve the parallelizability
of very large traces can be developed.
\end{enumerate}
In our experience complicated IBP reductions like massive 5-point
2-loop problems (e.g. \ref{subsec:pentabubble}) fit well under several
gigabytes, and thus are not limited by the trace sizes.

\section{Setting up}

To use \noun{Ratracer} first get its source code from its repository
over at

\hspace{1.5em}\url{https://github.com/magv/ratracer}

\noun{Ratracer} depends on \noun{FireFly}~\cite{Klappert:2019emp}
for rational function reconstruction, and on \noun{Flint}~\cite{Flint}
for modular arithmetics. These libraries will be automatically downloaded
when building \noun{Ratracer}; their dependencies \noun{Gmp}~\cite{GMP},
\noun{Mpfr}~\cite{MPFR}, and \noun{Jemalloc}~\cite{Jemalloc}
as well.

To download the dependent libraries, compile them, and build the \code{ratracer}
executable, run:

\begin{lstlisting}
$ make
\end{lstlisting}

\section{Using the command-line tool}

The command-line tool \code{ratracer} contains commands to solve
linear equation systems, operate on traces, and ultimately reconstruct
them. Its general usage pattern is:

\begin{lstlisting}
$ ratracer command args ... command args ... ...
\end{lstlisting}

The logic behind the commands is that the tool maintains its internal
state, and each command modifies it in a specific way. The internal
state includes:
\begin{itemize}
\item the current trace: the set of inputs, outputs, and the set of instructions;
the instructions come in two kinds:
\begin{itemize}
\item the low-level instructions (suitable for evaluation, but not for optimization),
\item the high-level instructions (suitable for optimization passes, but
not for evaluation);
\end{itemize}
\item the current set of linear equations and the list of integral families;
\item the current set of variable substitutions.
\end{itemize}

\subsection{List of provided commands}

\begin{itemize}
\item \manCm{load-trace} \manAr{file.trace}

Load the given trace. Automatically decompress the file
if the filename ends with \manQl{.gz}, \manQl{.bz2}, \manQl{.xz}, or \manQl{.zst}.

\item \manCm{save-trace} \manAr{file.trace}

Save the current trace to a file. Automatically compress
the file if the filename ends with \manQl{.gz}, \manQl{.bz2}, \manQl{.xz},
or \manQl{.zst}.

\item \manCm{show}

Show a short summary of the current trace.

\item \manCm{list-outputs} [\manFl{-{}-to}=\manAr{filename}]

Print the full list of outputs of the current trace.

\item \manCm{stat}

Collect and show the current code statistics.

\item \manCm{disasm} [\manFl{-{}-to}=\manAr{filename}]

Print a disassembly of the current trace.

\item \manCm{measure}

Measure the evaluation speed of the current trace.

\item \manCm{set} \manAr{name} \manAr{expression}

Set the given variable to the given expression in
the further traces created by \manCm{trace-expression},
\manCm{load-equations}, or loaded via \manCm{load-trace}.

\item \manCm{unset} \manAr{name}

Remove the mapping specified by \manCm{set}.

\item \manCm{load-trace} \manAr{file.trace}

Load the given trace.

\item \manCm{trace-expression} \manAr{filename}

Load a rational expression from a file and trace its
evaluation.

\item \manCm{keep-outputs} \manAr{filename}

Read a list of output name patterns from a file, one
pattern per line; keep all the outputs that match any
of these patterns, and erase all the others.

The pattern syntax is simple: \manQl{*} stands for any
sequence of characters, all other characters stand for
themselves.

\item \manCm{drop-outputs} \manAr{filename}

Read a list of output names from a file, one name per
line; erase all outputs contained in the list.

\item \manCm{optimize}

Optimize the current trace by propagating constants,
merging duplicate expressions, and erasing dead code.

\item \manCm{finalize}

Convert the (not yet finalized) code into a final low-level
representation that is smaller, and has drastically
lower memory usage. Automatically eliminate the dead
code while finalizing.

\item \manCm{unfinalize}

The reverse of \manCm{finalize} (i.e. convert low-level code
into high-level code), except that the eliminated code
is not brought back.

\item \manCm{reconstruct} [\manFl{-{}-to}=\manAr{filename}] [\manFl{-{}-threads}=\manAr{n}] [\manFl{-{}-factor-scan}] [\manFl{-{}-shift-scan}] [\manFl{-{}-bunches}=\manAr{n}] [\manFl{-{}-inmem}]

Reconstruct the rational form of the current trace using
the \noun{FireFly} library.

If the \manFl{-{}-inmem} flag is set, load the whole code
into memory during reconstruction; this increases the
performance especially with many threads, but comes at
the price of higher memory usage.

This command uses the \noun{FireFly} library for the reconstruction;
\manFl{-{}-factor-scan} and \manFl{-{}-shift-scan} flags enable
enable \noun{FireFly}'s factor scan and/or shift scan (which
are normally recommended); and \manFl{-{}-bunches} sets its
maximal bunch size.

\item \manCm{reconstruct0} [\manFl{-{}-to}=\manAr{filename}] [\manFl{-{}-threads}=\manAr{n}]

Same as \manCm{reconstruct}, but assumes that there are 0
input variables needed, and is therefore faster.

This command does not use the \noun{FireFly} library. The code
is always loaded into memory (as with the \manFl{-{}-inmem}
option of \manCm{reconstruct}).

\item \manCm{evaluate}

Evaluate the trace in terms of rational numbers.

Note that all the variables must have been previously
substitited, e.g. using the \manCm{set} command.

\item \manCm{define-family} \manAr{name} [\manFl{-{}-indices}=\manAr{n}]

Predefine an indexed family with the given number of
indices used in the equation parsing. This is only needed
to guarantee the ordering of the families, otherwise
they are auto-detected from the equation files.

\item \manCm{load-equations} \manAr{file.eqns}

Load linear equations from the given file in \noun{Kira} format,
tracing the expressions. Automatically decompress the file
if the filename ends with \manQl{.gz}, \manQl{.bz2}, \manQl{.xz},
or \manQl{.zst}.

\item \manCm{drop-equations}

Forget all current equations and families.

\item \manCm{solve-equations}

Solve all the currently loaded equations by gaussian
elimination, tracing the process.

Do not forget to \manCm{choose-equation-outputs} after this.

\item \manCm{choose-equation-outputs} [\manFl{-{}-family}=\manAr{name}] [\manFl{-{}-maxr}=\manAr{n}] [\manFl{-{}-maxs}=\manAr{n}] [\manFl{-{}-maxd}=\manAr{n}]

Mark the equations defining the specified integrals
as the outputs, so they could be later reconstructed.

That is, for each selected equation of the form
\manMa{- I_0 + \sum_i I_i C_i = 0}, add each of the coefficients
\manMa{C_i} as an output with the name \manMa{CO[I_0,I_i]}.

This command will fail if the equations are not in the
fully reduced form (i.e. after \manCm{solve-equations}).

The equations are filtered by the family name, maximal
sum of integral's positive powers (\manFl{-{}-maxr}), maximal
sum of negative powers (\manFl{-{}-maxs}), and/or maximal sum
of powers above one (\manFl{-{}-maxd}).

\item \manCm{show-equation-masters} [\manFl{-{}-family}=\manAr{name}] [\manFl{-{}-maxr}=\manAr{n}] [\manFl{-{}-maxs}=\manAr{n}] [\manFl{-{}-maxd}=\manAr{n}]

List the unreduced items of the equations filtered the
same as in \manCm{choose-equation-outputs}.

\item \manCm{dump-equations} [\manFl{-{}-to}=\manAr{filename}]

Dump the current list of equations with numeric coefficients.
This should only be needed for debugging.

\item \manCm{to-series} \manAr{varname} \manAr{maxorder}

Re-run the current trace treating each value as a series
in the given variable, and splitting each output into
separate outputs per term in the series.

The given variable is eliminated from the trace as a
result. The variable mapping is also reset.

\item \manCm{sh} \manAr{command}

Run the given shell command.

\item \manCm{help}

Show a help message and quit.

\end{itemize}

\subsection{Reconstructing a single expression}

To reconstruct a single arithmetic expression, prepare a file with
this expression:

\begin{lstlisting}
$ echo "1/(x+y) + 1/(x-y)" > expression.txt
\end{lstlisting}

Then trace the expression from the file, optimize the trace, and reconstruct
it:

\begin{lstlisting}
$ ratracer trace-expression expression.txt \
           optimize finalize \
           reconstruct --inmem --to=reconstruction.txt
[...]
$ cat reconstruction.txt
expression.txt =
  (x)/((-1/2)*y^2+1/2*x^2);
\end{lstlisting}

This tells us that
\begin{equation}
\frac{1}{x+y}+\frac{1}{x-y}=\frac{x}{\frac{1}{2}x^{2}-\frac{1}{2}y^{2}}.
\end{equation}

\subsubsection{Expanding an expression into a series}

Assuming that $x$ is small in the same expression as before, it is
possible to expand it into a series, keeping the first several orders
in $x$:

\begin{lstlisting}
$ ratracer trace-expression expression.txt \
           optimize finalize \
           to-series x 5 \
           optimize finalize \
           reconstruct --inmem --to=series.txt
$ cat series.txt
ORDER[expression.txt,x^1] =
  ((-2))/(y^2);
ORDER[expression.txt,x^2] =
  ((0))/(1);
ORDER[expression.txt,x^3] =
  ((-2))/(y^4);
ORDER[expression.txt,x^4] =
  ((0))/(1);
ORDER[expression.txt,x^5] =
  ((-2))/(y^6);
\end{lstlisting}

This gives us the series expansion of
\begin{equation}
\frac{1}{x+y}+\frac{1}{x-y}=-\frac{2}{y^{2}}x-\frac{2}{y^{4}}x^{3}-\frac{2}{y^{6}}x^{5}+\mathcal{O}(x^{6}).
\end{equation}

Note that it might be convenient to separately save the trace of \code{expression.txt},

\begin{lstlisting}
$ ratracer trace-expression expression.txt optimize finalize \
           save-trace expression.trace.gz
\end{lstlisting}

Here we have asked to automatically compress the trace with \code{gzip};
the recommended (and much faster) compressor however is \code{zstd},
which we can request using the file extension \code{.zst} (if the
\code{zstd} program is installed).

In any case, the obtained trace can now be either restored fully as 

\begin{lstlisting}
$ ratracer load-trace expression.trace.gz reconstruct
\end{lstlisting}

Or expanded into series and reconstructed

\begin{lstlisting}
$ ratracer load-trace expression.trace.gz \
           to-series x 5 optimize finalize reconstruct
\end{lstlisting}

Of course, it is also possible to similarly save the trace after the
series expansion.

\subsection{Solving IBP relations (together with \protect\noun{Kira})}

\begin{figure}[H]
\begin{centering}
\raisebox{0.5ex}{\scalebox{1.00}{\begin{tikzpicture}
	\begin{pgfonlayer}{nodelayer}
		\node [style=none] (0) at (-2, 0) {};
		\node [style=dot] (1) at (0, 1.25) {};
		\node [style=dot] (2) at (0, -1.25) {};
		\node [style=none] (3) at (2, 0) {};
		\node [style=dot] (4) at (-1.25, 0) {};
		\node [style=dot] (5) at (1.25, 0) {};
		\node [style=none] (11) at (-2.25, 0) {$q$};
		\node [style=none] (12) at (2.25, 0) {$q$};
		\node [style=none, rotate=45] (13) at (-0.75, 1) {$l_1,m_1$};
		\node [style=none, rotate=-45] (14) at (0.75, 1) {$l_2,m_1$};
		\node [style=none, rotate=-45] (15) at (-0.75, -1) {$q+l_1,m_2$};
		\node [style=none, rotate=45] (16) at (0.75, -1) {$q+l_2,m_2$};
		\node [style=none, rotate=90] (17) at (-0.25, 0) {$l_1-l_2$};
	\end{pgfonlayer}
	\begin{pgfonlayer}{edgelayer}
		\draw [style=massive1 edge, arrow] (0.center) to (4);
		\draw [style=massive1 edge, arrow] (5) to (3.center);
		\draw [style=edge, arrow] (2) to (1);
		\draw [style=massive2 edge, arrow] (1) to (4);
		\draw [style=massive2 edge, arrow] (5) to (1);
		\draw [style=massive3 edge, arrow] (2) to (5);
		\draw [style=massive3 edge, arrow] (4) to (2);
	\end{pgfonlayer}
\end{tikzpicture}}}
\par\end{centering}
\caption{\label{fig:diamond2l-detailed}Massive 2-loop diamond with 2 masses,
2 massive legs, 5 propagators, and the total of 2 scaleless ratios.}
\end{figure}
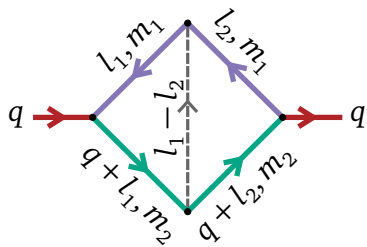

\noun{Ratracer} can import linear equations in \noun{Kira} format
and solve them, so any IBP reduction \noun{Kira} allows for can also
be performed in combination with \noun{Ratracer} for better performance
and flexibility.

Let us illustrate this by reducing integrals in the two-loop diamond
topology of \ref{fig:diamond2l-detailed}. First start by preparing
the \noun{Kira} configuration:

\begin{lstlisting}
$ mkdir config
$ cat >config/integralfamilies.yaml <<EOF
integralfamilies:
  - name: "diamond"
    loop_momenta: ["l1", "l2"]
    top_level_sectors: [b11111]
    propagators:
      - ["l1 - l2", "0"]
      - ["l1", "m1sq"]
      - ["l2", "m1sq"]
      - ["l1 + q", "m2sq"]
      - ["l2 + q", "m2sq"]
EOF

$ cat >config/kinematics.yaml <<EOF
kinematics:
  incoming_momenta: ["q"]
  outgoing_momenta: []
  kinematic_invariants:
    - ["s", 2]
    - ["m1sq", 2]
    - ["m2sq", 2]
  scalarproduct_rules:
    - [["q", "q"], "s"]
  symbol_to_replace_by_one: "s"
EOF
\end{lstlisting}
Next, prepare the \noun{Kira} job specification, e.g. requesting
the export for all equations containing integrals with up to 3 numerators
and up to 2 dots, and run it:

\begin{lstlisting}
$ cat >export-equations.yaml <<EOF
jobs:
  - reduce_sectors:
      reduce:
        - { topologies: [diamond], sectors: [b11111], r:8, s:3 }
      select_integrals:
        select_mandatory_recursively:
          - { topologies: [diamond], sectors: [b11111], r:8, s:3, d:2 }
      run_symmetries: true
      run_initiate: input
      run_triangular: false
      run_back_substitution: false
      run_firefly: false
EOF

$ kira export-equations.yaml
\end{lstlisting}

The equations will be exported to a set of \code{input\_kira/diamond/SYSTEM\_{*}.kira.gz}
files (compressed with \code{gzip}). We can now load them into \noun{Ratracer}
and solve them (note that \noun{Ratracer} automatically handles compressed
equation files):

\begin{lstlisting}
$ ratracer \
    $(for x in input_kira/*/*kira.gz; do echo load-equations $x; done) \
    solve-equations \
    choose-equation-outputs --family=diamond --maxr=8 --maxs=3 --maxd=2 \
    drop-equations \
    optimize finalize \
    reconstruct --inmem --threads=8 --factor-scan --shift-scan \
        --to=ibp-tables.txt
\end{lstlisting}

The solution file will look like this:

\begin{lstlisting}
$ cat ibp-tables.txt
CO[diamond[1,1,1,1,3],diamond[0,1,1,1,1]] =
  ((-16/3)*m1sq*d^2+(-7/3)*d^2*m2sq^2+1/3*m1sq*d^3+(-7/3)*d^2+...
CO[diamond[1,1,1,1,3],diamond[1,-1,1,1,0]] =
  ((-5147/40)*m1sq^3*d^2*m2sq^3+1/6*d^4*m2sq^6+37/32*m1sq*d^4*...
...
\end{lstlisting}

Here \code{CO{[}diamond{[}1,1,1,1,3{]},diamond{[}0,1,1,1,1{]}{]}}
stands for ``coefficient of the integral $\text{diamond}_{1,1,1,1,3}$
w.r.t. the master integral $\text{diamond}_{0,1,1,1,1}$'', so
\begin{equation}
\begin{aligned}\text{diamond}_{1,1,1,1,3} & =\left(-\frac{16}{3}m_{1}^{2}d^{2}-\frac{7}{3}d^{2}m_{2}^{4}+\frac{1}{3}m_{1}^{2}d^{3}-\frac{7}{3}d^{2}+\dots\right)\text{diamond}_{0,1,1,1,1}+\\
 & +\left(-\frac{5147}{40}m_{1}^{6}d^{2}m_{2}^{6}+\frac{1}{6}d^{4}m_{2}^{12}+\frac{37}{32}m_{1}^{2}d^{4}\times\dots\right)\text{diamond}_{1,-1,1,1,0}+\\
 & +\dots.
\end{aligned}
\end{equation}

\subsubsection{Specifying preferred master integrals}

By default \noun{Ratracer} will choose some integrals as masters,
but selecting a better master basis can turn a hard problem into an
easy one. Of particular note are the $d$-factorizing bases~\cite{Smirnov:2020quc,Usovitsch:2020jrk},
where the dependence of the reduction coefficients on the dimensional
parameter $d$ in the denominator is factorized, which greatly simplifies
the reduction process.

To force \noun{Ratracer} to choose a particular basis one can exploit
the fact that \noun{Ratracer} maintains a strict order of integrals
when solving an equation system: the integrals of the families defined
first are eliminated last. Thus, defining an auxiliary family, e.g.
named ``master'', and adding equations setting $\text{master}_{i}$
to the desired integral (or a linear combination of integrals), will
make sure that these integrals will be eliminated last, becoming the
preferred master integrals.

To achieve this, first create a file with the additional equations:

\begin{lstlisting}
$ cat >master-equations.txt <EOF
master@1*1
diamond[0,1,1,0,0]*-1

master@2*1
diamond[0,0,1,1,0]*-1

master@3*1
diamond[1,0,1,1,0]*-1

master@4*1
diamond[1,-1,1,1,0]*-1

master@5*1
diamond[1,0,1,1,-1]*-1

master@6*1
diamond[0,1,1,1,0]*-1

master@7*1
diamond[0,0,0,1,1]*-1

master@8*1
diamond[0,1,0,1,1]*-1

master@9*1
diamond[0,1,1,1,1]*-1
EOF
\end{lstlisting}

Then, solve the equations as before, but with additionally defining
the \code{master} integral family (to make sure it is defined first),
and loading this equation file in addition to what \noun{Kira} has
exported:

\begin{lstlisting}
$ ratracer \
    define-family master \
    load-equations master-equations.txt \
    $(for x in input_kira/*/*kira.gz; do echo load-equations $x; done) \
    solve-equations \
    choose-equation-outputs --family=diamond --maxr=8 --maxs=3 --maxd=2 \
    drop-equations \
    optimize finalize \
    reconstruct --inmem --threads=8 --factor-scan --shift-scan \
        --to=ibp-tables.txt
\end{lstlisting}

This time the result will look like the following:

\begin{lstlisting}
$ cat ibp-tables.txt
CO[diamond[1,1,1,1,3],master@9] =
  ((-16/3)*m1sq*d^2+(-7/3)*d^2*m2sq^2+1/3*m1sq*d^3+(-7/3)*d^2+...
...
\end{lstlisting}

Naturally, specifying more master integrals than there are, or less
than there should be is harmless, because all we are doing is modifying
the priority of the elimination.

\subsubsection{Master-wise and sector-wise reconstruction}

Instead of reconstructing the equation system solutions right away
one can first operate on the trace. Let us start by saving it to a
file.

\begin{lstlisting}
$ ratracer \
    $(for x in input_kira/*/*kira.gz; do echo load-equations $x; done) \
    solve-equations \
    choose-equation-outputs --family=diamond --maxr=8 --maxs=3 --maxd=2 \
    optimize finalize \
    save-trace trace.zst
\end{lstlisting}

Now let us select only the outputs with the first master integral
($\text{diamond}_{0,1,1,0,0}$).

\begin{lstlisting}
$ cat >output-list.txt <<EOF
CO[*,diamond[0,1,1,0,0]]
EOF
$ ratracer \
    load-trace trace.zst \
    keep-outputs output-list.txt \
    unfinalize finalize \
    save-trace trace-master1.zst
\end{lstlisting}

Now \code{trace-master1.zst} only contains the coefficients with
the first master, and all the remaining computations are removed from
it, and it is therefore faster to evaluate (by a factor of 3.6 in
this case):

\begin{lstlisting}
$ ratracer load-trace trace.zst measure
...
Average time: 0.0001809s after 4095 evals 
...
$ ratracer load-trace trace-master1.zst measure
...
Average time: 4.959e-05s after 16383 evals 
...
\end{lstlisting}

By saving traces of coefficients with each of the masters into separate
files and reconstructing them separately we can achieve ``master-wise''
reduction. In other systems the same could be achieved by setting
other master integrals to zero and re-running the IBP solution; with
\noun{Ratracer} we only need to select the appropriate outputs, and
drop the computations that don't contribute to them.

Similarly, we can split the computation by the integral being reduced:

\begin{lstlisting}
$ cat >output-list.txt <<EOF
CO[diamond[1,1,1,1,3],*]
EOF
$ ratracer \
    load-trace trace.zst \
    keep-outputs output-list.txt \
    unfinalize finalize \
    save-trace trace-integral1.zst
\end{lstlisting}

The resulting trace is, as expected, faster than the original one
(by a factor of 9.3x this time):

\begin{lstlisting}
$ ratracer load-trace trace-integral1.zst measure
...
Average time: 1.955e-05s after 32767 evals 
...
\end{lstlisting}

By selectively choosing which integrals should go into which files,
we can achieve sector-wise reduction.

Finally, we don't need to stop at splitting by master or by integral,
we can do both and have a separate trace for each coefficient, or
any subset of them if we wish.

\subsubsection{Reconstruction with kinematics set to constants}

With \noun{Ratracer} it is possible to set variables to some expressions
(e.g. constants) before loading a trace, so that the trace would use
that as its input. One use of this functionality is to evaluate the
IBP reduction coefficients at particular fixed values of the kinematic
invariants. This can be done the following way:

\begin{lstlisting}
$ ratracer \
    set m1sq 3 \
    set m2sq 5/2 \
    load-trace trace.zst \
    reconstruct
...
CO[diamond[1,1,1,1,3],diamond[0,1,1,1,1]] =
  ((-8/19773)*d^3+464/59319*d^2+(-2536/59319)*d+1360/19773)/((-1/6)*d+1);
...
\end{lstlisting}

Note that the reconstruction is performed in only one variable: $d$,
because all the other variables are now fixed.

Another use of this feature is to perform reconstruction on a particular
line (or, more generally, hypersurface) in the kinematic parameter
space. For example, we can set $m_{1}^{2}=3/5\lambda$ and $m_{2}^{2}=7/11\lambda$,
and then reconstruct only in $\lambda$ (and $d$):

\begin{lstlisting}
$ ratracer \
    set m1sq '3/5*lambda' \
    set m2sq '7/11*lambda' \
    load-trace trace.zst \
    reconstruct
...
CO[diamond[1,1,1,1,3],diamond[0,1,1,1,1]] =
  ((-15)+(-38/165)*lambda*d^2+1/6*d^3+21/2*d+(-12/605)*lambda^2+...
...
\end{lstlisting}

This mode of usage can be useful for finding the values of the integrals
via differential equations along particular lines as is done in~\cite{Hidding:2020ytt,Liu:2022chg,Armadillo:2022ugh},
if the full reduction is too complicated.

\subsubsection{Reconstructing a truncated series in \texorpdfstring{$\varepsilon$}{epsilon}}

A major increase in reconstruction performance can be obtained by
noting that fairly often the dimensional parameter~$d$ enters the
reconstructed expressions at a fairly high power (as $d^{6}$ in the
case of \ref{fig:diamond2l-detailed}), but in practice most of this
information is not needed: what is needed is the expansion of the
coefficients in $\varepsilon$ (with $d=4-2\epsilon$), and only the
first few terms of it, up to $\mathcal{O}(\epsilon^{1})$ or $\mathcal{O}(\epsilon^{0})$
usually, so restoring the dependence on $d^{6}$ (i.e. $\varepsilon^{6}$)
is a waste of time. Fortunately, we can avoid this waste by expanding
the coefficients into a series in~$\varepsilon$ \emph{before} the
reconstruction, and then only reconstruct the required orders.

Here let us expand up to $\mathcal{O}(\varepsilon^{0})$ terms:

\begin{lstlisting}
$ ratracer \
    set d '4-2*eps' \
    load-trace trace.zst \
    to-series eps 1 \
    optimize finalize \
    reconstruct
...
ORDER[CO[diamond[1,1,1,1,3],diamond[0,1,1,1,1]],eps^0] =
  (1+4*m1sq+m2sq^2+(-2)*m2sq+m1sq^2+(-2)*m1sq*m2sq)/(15*m1sq^4*...
ORDER[CO[diamond[1,1,1,1,3],diamond[0,1,1,1,1]],eps^1] =
  ((-2)*m1sq)/(15*m1sq^4*m2sq^2+15*m2sq^2+15*m1sq^2+15*m2sq^4+...
ORDER[CO[diamond[1,1,1,1,3],diamond[1,-1,1,1,0]],eps^-1] =
  (6*m1sq^3*m2sq+3/2*m1sq*m2sq^4+9*m1sq^3*m2sq^2+(-6)*m1sq^4*m2sq+...
...
\end{lstlisting}

The performance effect of this expansion is two-fold:
\begin{itemize}
\item First, the variable in which the expansion is made is removed from
the reconstruction, making it faster (fewer probes are required).
\item Second, more outputs are needed and their per-probe evaluation time
might be larger than that of the original problem.
\end{itemize}
In practice this works out so that expanding to $\mathcal{O}(\varepsilon^{0})$
improves the overall reconstruction time by a factor of 2--5; less
so for $\mathcal{O}(\varepsilon^{1})$, and if a high enough expansion
order is requested, then there is no performance benefit. For this
reason it is important to specify the expansion order as low as is
practically needed. For example, if $CO[\text{diamond}_{1,1,1,1,3},\text{diamond}_{0,1,1,1,1}]$
is only needed to $\mathcal{O}(\varepsilon^{0}$), but the other coefficients
are needed to $\mathcal{O}(\varepsilon^{1}$), we can expand everything
up to $\mathcal{O}(\varepsilon^{1}$) and then remove the coefficients
we don't need:

\begin{lstlisting}
$ cat >droplist.txt <<EOF
ORDER[CO[diamond[1,1,1,1,3],diamond[0,1,1,1,1]],eps^1]
EOF
$ ratracer \
    set d '4-2*eps' \
    load-trace trace.zst \
    to-series eps 1 \
    drop-outputs droplist.txt \
    optimize finalize \
    reconstruct
...
ORDER[CO[diamond[1,1,1,1,3],diamond[0,1,1,1,1]],eps^0] =
  (1+4*m1sq+m2sq^2+(-2)*m2sq+m1sq^2+(-2)*m1sq*m2sq)/(15*m1sq^4*...
ORDER[CO[diamond[1,1,1,1,3],diamond[1,-1,1,1,0]],eps^-1] =
  (6*m1sq^3*m2sq+3/2*m1sq*m2sq^4+9*m1sq^3*m2sq^2+(-6)*m1sq^4*m2sq+...
...
\end{lstlisting}

Speaking more generally, the series does not need to be in~$\varepsilon$,
and any variable can be used for the expansion. For example, if one
is interested in some particular high-energy region, expansion in
a kinematic invariant may be appropriate. Multiple expansions are
possible too.

\section{Using the C++ library}

In addition to the command-line tool \noun{Ratracer} provides a library
\code{ratracer.h}, which allows users to construct traces of arbitrary
computations defined as C++ code. The command-line tool itself is
built on top of this library. A program using it first needs to make
sure the \noun{Flint}~\cite{Flint} and \noun{Gmp}~\cite{GMP}
libraries are installed, then include the header file as

\begin{lstlisting}
#include "ratracer.h"
\end{lstlisting}

and link the resulting program together with the \noun{Flint} and
\noun{Gmp} libraries, for example:

\begin{lstlisting}
c++ -o custom_program custom_program.cpp -lflint -lgmp
\end{lstlisting}

\subsection{List of provided functions}

The interface provided by \code{ratracer.h} consist of an opaque
structure representing a value (a rational modulo a prime):

\begin{lstlisting}[language={C++}]
struct Value { };
\end{lstlisting}

and a class that contains all the arithmetic operations one can perform
on these values:

\begin{lstlisting}[language={C++}]
struct Tracer {
    void clear();
    size_t checkpoint();
    void rollback(size_t checkpoint);
    Value var(size_t idx);
    Value of_int(int64_t x);
    Value of_fmpz(const fmpz_t x);
    bool is_zero(const Value &a);
    bool is_minus1(const Value &a);
    Value mul(const Value &a, const Value &b);
    Value mulint(const Value &a, int64_t b);
    Value add(const Value &a, const Value &b);
    Value addint(const Value &a, int64_t b);
    Value sub(const Value &a, const Value &b);
    Value addmul(const Value &a,
                 const Value &b1,
                 const Value &b2);
    Value inv(const Value &a);
    Value neginv(const Value &a);
    Value neg(const Value &a);
    Value pow(const Value &base, long exp);
    Value shoup_precomp(const Value &a);
    Value shoup_mul(const Value &a,
                    const Value &aprecomp,
                    const Value &b);
    Value div(const Value &a, const Value &b);
    void assert_int(const Value &a, int64_t n);
    void add_output(const Value &src, const char *name);
    size_t input(const char *name, size_t len);
    size_t input(const char *name);
    int save(const char *path);
};
\end{lstlisting}

The \code{Tracer} structure should be initialized by:

\begin{lstlisting}[language={C++}]
Tracer tracer_init();
\end{lstlisting}

\subsection{An example}

A very simple usage example would be this fragment that records a
trace with
\begin{equation}
\mathrm{expr}=x^{2}+3y
\end{equation}

\begin{lstlisting}[language={C++}]
#include "ratracer.h"
int main() {
    Tracer tr = tracer_init();
    Value x = tr.var(tr.input("x"));
    Value y = tr.var(tr.input("y"));
    Value expr = tr.add(tr.pow(x, 2), tr.mulint(y, 3));
    tr.add_output(expr, "expr");
    tr.save("example.trace.gz");
    return 0;
}
\end{lstlisting}

To compile and run the program on a normal Unix machine, execute:

\begin{lstlisting}
$ c++ -o example example.cpp -lflint -lgmp
$ ./example
\end{lstlisting}

The resulting trace \code{example.trace.gz} can be examined with
\code{ratracer}:

\begin{lstlisting}
$ ratracer load-trace example.trace.gz show
[...]
0.0002 +0.0001 * load-trace
0.0002 +0.0000   Importing 'example.trace.gz'
0.0055 +0.0053 * show
0.0055 +0.0000   Current trace:
0.0055 +0.0000   - inputs: 2
0.0055 +0.0000     0) x
0.0055 +0.0000     1) y
0.0055 +0.0000   - outputs: 1
0.0055 +0.0000     0) expr
0.0055 +0.0000   - big integers: 0
0.0055 +0.0000   - instructions: 0B final, 16kB temp
0.0055 +0.0000   - locations: 0 final (0B), 1024 temp (8.00kB)
0.0055 +0.0000   - next free location: 1024
0.0055 +0.0000   Current equation set:
0.0055 +0.0000   - families: 0
0.0055 +0.0000   - equations: 0
0.0055 +0.0000   Active variable replacements: none
0.0055 +0.0000   Runtime: 0.000s user time,
                          0.003s system time,
                          6.40MB maximum RSS
\end{lstlisting}

It can also be disassembled into human-readable format:

\begin{lstlisting}
$ ratracer load-trace example.trace.gz disasm
[...]
0.0055 +0.0000 * disasm
# ninputs = 2
# noutputs = 1 
# nconstants = 0 
# nfinlocations = 0
# nlocations = 1024
# low-level code (0B)
# high-level code (16384B)
0 = var #0 'x'
1 = var #1 'y'
2 = int #3
3 = mul 1 2
4 = mul 0 0
5 = add 4 3
6 = output 5 #0 'expr'
\end{lstlisting}

The resulting expression can be reconstructed too. For best results,
we first optimize the trace, and perform the reconstruction in memory:

\begin{lstlisting}
$ ratracer load-trace example.trace.gz \
           optimize \
           finalize \
           reconstruct --inmem
[...]
expr =
  (x^2+3*y)/(1);
\end{lstlisting}

\section{Benchmarks}

To put the performance of \noun{Ratracer} into context, we have compared
it with two state-of-the-art IBP solvers: \noun{Kira} and \noun{Fire6}.
Specifically, we have obtained IBP reduction tables for several example
topologies (see \ref{fig:tth2l_b25}, \ref{fig:xbox2l2m}, \ref{fig:tth2l_b16},
\ref{fig:diamond3l}, and \ref{fig:box2l} from \ref{app:benchmark-tables})
using the following methods:
\begin{itemize}
\item ``\noun{Ratracer}'': using \noun{Kira}\footnote{\noun{Kira} version 2.2 as of 2022-05-04, built with the default
options plus the \noun{Jemalloc}~\cite{Jemalloc} memory allocator.} to generate the equations, then using \noun{Ratracer} to solve them,
and finally using \noun{Ratracer} and \noun{FireFly}\footnote{\noun{FireFly} version 2.0.3 as of 2022-05-30, built with the default
options.} to reconstruct the answer.
\item ``\noun{Ratracer}+$\mathcal{O}(\varepsilon^{n})$'': same as the
previous method, but additionally asking \noun{Ratracer} to expand
the results in~$\varepsilon$ up to (and including) terms of the
order $\varepsilon^{n}$ before proceeding with the reconstruction.
\item ``\noun{Ratracer}+Scan'': same as \noun{Ratracer}, but with the
``shift scan'' and ``factor scan'' options of \noun{FireFly}
enabled.
\item ``\noun{Kira}+\noun{FireFly}'': using \noun{Kira} together with
its \noun{FireFly} backend. Following~\cite{Klappert:2020nbg} we
set the ``maximal bunch size'' parameter to~4 (\noun{Ratracer}
uses the same value).
\item ``\noun{Kira}'': using \noun{Kira} in its default configuration
(i.e. using \noun{Fermat}~\cite{Fermat} for symbolic expression
simplification). This method does not directly use modular arithmetics
(aside from a preparation step, and possibly inside \noun{Fermat}),
so we only provide these measurements as a reference.
\item ``\noun{Fire6}'': using \noun{Fire6}\footnote{\noun{Fire6} version 6.4.2, as of 2022-02-29, build with the default
options.} and \noun{LiteRed}~\cite{Lee:2013mka} as described in~\cite{Smirnov:2019qkx}.
This method also does not use modular arithmetics (aside from a preparation
step), so we only provide these measurements as a reference.
\item ``\noun{Fire6}+Hints'': same as \noun{Fire6}, but with with a
separate step preparing the ``hints file'' as recommended in~\cite{Smirnov:2019qkx}.
\end{itemize}
Among the benchmarks we consider two to be realistic ``hard'' problems:
the two-loop massive hexatriangle described in \ref{subsec:hexatriangle}
and the two-loop massive non-planar box from \ref{subsec:xbox2l2m}.
The remaining ones are provided for comparison.

\subsection{What is measured?}

In all cases the performance is measured on the same machine running
openSUSE~15.3 with an AMD EPYC~7282 processor, an SSD disk, and
sufficient RAM. The programs are asked to use 8~threads, and hyperthreading
is effectively disabled via a Linux thread CPU affinity setting.\footnote{Using \noun{Hypothread}, \url{https://github.com/magv/hypothread}.}
The code for running the benchmarks has is publicly available on GitHub.\footnote{See \url{https://github.com/magv/ibp-benchmark}.}

The detailed benchmark result tables can be found in \ref{app:benchmark-tables}.
The reported performance figures are:
\begin{itemize}
\item ``Total time'': time from start to finish (from the plain configuration
files to the final IBP substitution tables), including any preparation
time that the method requires.
\item ``\noun{FireFly} time'': pure reduction time, excluding any preparation.
\item ``Probes'': number of black-box probes \noun{FireFly} has evaluated
to get the final reconstruction.
\item ``Probe time'': average time of a single black-box probe, as reported
by \noun{FireFly}.
\item ``\noun{FireFly} eff.'': the efficiency of \noun{FireFly}'s rational
reconstruction implementation, measured as the ratio of the total
probe time to the total \noun{FireFly} time scaled by the thread
count.
\item ``Memory'': peak memory usage during the reduction.
\item ``Disk'': peak disk usage during the reduction.
\end{itemize}

\subsection{Benchmark summary}

Here is a short summary of \noun{Ratracer} performance in all the
benchmarks from \ref{app:benchmark-tables}:
\begin{center}
\begin{tabular}{ccccccc}
\toprule 
 &  & Probe time & Total time & Total time & $\mathcal{O}(\varepsilon^{0})$ & $\mathcal{O}(\varepsilon^{1})$\tabularnewline
 &  & speedup & speedup vs. & speedup & speedup & speedup\tabularnewline
 &  & vs. \noun{Kira} & \noun{Kira}+\noun{FireFly} & vs. \noun{Kira} &  & \tabularnewline
\midrule
\midrule 
\ref{subsec:hexatriangle} & \raisebox{0.5ex}{\scalebox{0.50}{\begin{tikzpicture}
	\begin{pgfonlayer}{nodelayer}
		\node [style=dot] (0) at (-0.75, -0.5) {};
		\node [style=dot] (1) at (0, -1) {};
		\node [style=dot] (2) at (0.75, -0.5) {};
		\node [style=dot] (3) at (0.75, 0.5) {};
		\node [style=dot] (4) at (0, 1) {};
		\node [style=dot] (5) at (-0.75, 0.5) {};
		\node [style=dot] (6) at (1.5, 0) {};
		\node [style=none] (7) at (-1.5, -0.5) {};
		\node [style=none] (8) at (0, -1.75) {};
		\node [style=none] (9) at (2.25, 0) {};
		\node [style=none] (10) at (-1.5, 0.5) {};
		\node [style=none] (11) at (0, 1.75) {};
		\node [style=none] (12) at (-0.75, 1) {$m_1$};
		\node [style=none] (13) at (1, 0) {$m_1$};
		\node [style=none] (14) at (2, 0.25) {$m_2$};
	\end{pgfonlayer}
	\begin{pgfonlayer}{edgelayer}
		\draw [style=edge] (8.center) to (1);
		\draw [style=edge] (7.center) to (0);
		\draw [style=edge] (5) to (0);
		\draw [style=edge] (0) to (1);
		\draw [style=edge] (1) to (2);
		\draw [style=edge] (3) to (4);
		\draw [style=massive2 edge] (9.center) to (6);
		\draw [style=massive1 edge] (10.center) to (5);
		\draw [style=massive1 edge] (5) to (4);
		\draw [style=massive1 edge] (4) to (11.center);
		\draw [style=massive1 edge] (2) to (3);
		\draw [style=massive1 edge] (3) to (6);
		\draw [style=massive1 edge] (6) to (2);
	\end{pgfonlayer}
\end{tikzpicture}}} & 20 & 5.2 & 1.2 & 3.2 & 2.4\tabularnewline
\midrule 
\ref{subsec:xbox2l2m} & \raisebox{0.5ex}{\scalebox{0.50}{\begin{tikzpicture}
	\begin{pgfonlayer}{nodelayer}
		\node [style=none] (0) at (-1.5, -0.75) {};
		\node [style=none] (1) at (-1.5, 0.75) {};
		\node [style=dot] (2) at (-0.75, -0.75) {};
		\node [style=dot] (3) at (-0.75, 0.75) {};
		\node [style=dot] (4) at (0, -0.75) {};
		\node [style=dot] (5) at (0, 0.75) {};
		\node [style=dot] (6) at (0.75, -0.75) {};
		\node [style=dot] (7) at (0.75, 0.75) {};
		\node [style=none] (8) at (1.5, -0.75) {};
		\node [style=none] (9) at (1.5, 0.75) {};
		\node [style=none] (10) at (1.25, 0.5) {$m_1$};
		\node [style=none] (11) at (-1, 0.25) {$m_2$};
	\end{pgfonlayer}
	\begin{pgfonlayer}{edgelayer}
		\draw [style=edge] (1.center) to (3);
		\draw [style=edge] (0.center) to (2);
		\draw [style=edge] (2) to (4);
		\draw [style=edge] (2) to (5);
		\draw [style=edge] (4) to (6);
		\draw [style=edge] (6) to (7);
		\draw [style=edge] (6) to (8.center);
		\draw [style=massive1 edge] (3) to (5);
		\draw [style=massive1 edge] (5) to (7);
		\draw [style=massive1 edge] (7) to (9.center);
		\draw [style=massive2 edge] (3) to (4);
	\end{pgfonlayer}
\end{tikzpicture}}} & 7.8 & 6.0 & 37 & 2.7 & 1.4\tabularnewline
\midrule 
\ref{subsec:pentabubble} & \raisebox{0.5ex}{\scalebox{0.50}{\begin{tikzpicture}
	\begin{pgfonlayer}{nodelayer}
		\node [style=none] (0) at (-1.25, 0.5) {};
		\node [style=dot] (1) at (-0.5, 0.5) {};
		\node [style=none] (2) at (1, 1.25) {};
		\node [style=dot] (3) at (0.5, 0.75) {};
		\node [style=dot] (4) at (-0.5, -0.5) {};
		\node [style=none] (5) at (-1.25, -0.5) {};
		\node [style=dot] (6) at (0.5, -0.75) {};
		\node [style=none] (7) at (1, -1.25) {};
		\node [style=dot] (8) at (1, 0) {};
		\node [style=none] (9) at (1.75, 0) {};
		\node [style=none] (10) at (2.25, 0) {$m_2$};
		\node [style=none] (11) at (1.25, -1) {$m_1$};
		\node [style=none] (12) at (1.25, 1) {$m_1$};
	\end{pgfonlayer}
	\begin{pgfonlayer}{edgelayer}
		\draw [style=edge] (1) to (0.center);
		\draw [style=massive1 edge] (3) to (2.center);
		\draw [style=massive2 edge] (8) to (9.center);
		\draw [style=massive1 edge] (6) to (7.center);
		\draw [style=edge] (4) to (5.center);
		\draw [style=massive1 edge, bend right] (1) to (3);
		\draw [style=massive1 edge, bend left=330] (3) to (1);
		\draw [style=massive1 edge] (3) to (8);
		\draw [style=massive1 edge] (8) to (6);
		\draw [style=edge] (1) to (4);
		\draw [style=edge] (4) to (6);
	\end{pgfonlayer}
\end{tikzpicture}}} & 8.9 & 2.1 & 1/5.1 & 5.5 & 4.0\tabularnewline
\midrule 
\ref{subsec:diamond3l} & \raisebox{0.5ex}{\scalebox{0.50}{\begin{tikzpicture}
	\begin{pgfonlayer}{nodelayer}
		\node [style=dot] (0) at (-1, 0) {};
		\node [style=dot] (1) at (-0.25, 0.75) {};
		\node [style=dot] (2) at (-0.25, -0.75) {};
		\node [style=dot] (5) at (0.5, 0.75) {};
		\node [style=dot] (6) at (0.5, -0.75) {};
		\node [style=dot] (7) at (1.25, 0) {};
		\node [style=none] (8) at (-1.75, 0) {};
		\node [style=none] (9) at (2, 0) {};
		\node [style=none] (10) at (-1.5, 0.25) {$s$};
		\node [style=none] (12) at (-0.75, 0.75) {$m_1$};
		\node [style=none] (13) at (1, -0.75) {$m_2$};
	\end{pgfonlayer}
	\begin{pgfonlayer}{edgelayer}
		\draw [style=massive2 edge] (5) to (7);
		\draw [style=massive3 edge] (7) to (6);
		\draw [style=massive3 edge] (2) to (0);
		\draw [style=massive2 edge] (0) to (1);
		\draw [style=massive1 edge] (0) to (8.center);
		\draw [style=massive1 edge] (7) to (9.center);
		\draw [style=edge] (1) to (2);
		\draw [style=edge] (5) to (6);
		\draw [style=massive2 edge] (1) to (5);
		\draw [style=massive3 edge] (2) to (6);
	\end{pgfonlayer}
\end{tikzpicture}}} & 26 & 1.7 & 1/3.3 & 2.3 & 1.7\tabularnewline
\midrule 
\ref{subsec:box2l} & \raisebox{0.5ex}{\scalebox{0.50}{\begin{tikzpicture}
	\begin{pgfonlayer}{nodelayer}
		\node [style=none] (0) at (-1.5, 0.75) {};
		\node [style=none] (1) at (-1.5, -0.75) {};
		\node [style=dot] (2) at (-0.75, 0.75) {};
		\node [style=dot] (3) at (-0.75, -0.75) {};
		\node [style=dot] (4) at (0, 0.75) {};
		\node [style=dot] (5) at (0, -0.75) {};
		\node [style=dot] (6) at (0.75, 0.75) {};
		\node [style=dot] (7) at (0.75, -0.75) {};
		\node [style=none] (8) at (1.5, 0.75) {};
		\node [style=none] (9) at (1.5, -0.75) {};
		\node [style=none] (10) at (1, 0) {$m$};
	\end{pgfonlayer}
	\begin{pgfonlayer}{edgelayer}
		\draw [style=edge] (0.center) to (2);
		\draw [style=edge] (1.center) to (3);
		\draw [style=edge] (6) to (8.center);
		\draw [style=edge] (7) to (9.center);
		\draw [style=massive1 edge] (2) to (4);
		\draw [style=massive1 edge] (4) to (6);
		\draw [style=massive1 edge] (6) to (7);
		\draw [style=massive1 edge] (7) to (5);
		\draw [style=massive1 edge] (5) to (3);
		\draw [style=massive1 edge] (3) to (2);
		\draw [style=massive1 edge] (4) to (5);
	\end{pgfonlayer}
\end{tikzpicture}}} & 9.6 & 5.2 & 2.6 & 4.3 & 2.3\tabularnewline
\bottomrule
\end{tabular}
\par\end{center}

To summarize the benchmark results:
\begin{itemize}
\item \noun{Ratracer} consistently improves the probe time by a factor
of 3--20 compared to \noun{Kira}.
\item At the same time the overall runtime is improved by a factor of 5
for the ``hard'' examples compared to \noun{Kira}+\noun{FireFly}.
This is not always as big as the probe time improvement would suggest;
the reason is that with probes this fast the overhead of the rational
reconstruction in \noun{FireFly} begins to dominate. In fact, on
some examples like \ref{subsec:diamond3l} \noun{Ratracer} probes
are so fast that \noun{FireFly} efficiency drops below~3\%.
\item Interestingly, plain \noun{Kira} (i.e. \noun{Kira}+\noun{Fermat})
is normally faster than \noun{Kira}+\noun{FireFly}, with the exception
of the non-planar box example (\ref{subsec:xbox2l2m}), where it spends
a big part of its reduction time waiting for a single \noun{Fermat}
process to finish.
\item A speedup by a factor of 2.5--5.5 is possible if one expands the
reduction coefficients to~$\mathcal{O}(\varepsilon^{0})$, or only
1.4--4 if $\mathcal{O}(\varepsilon^{1})$ is needed.
\end{itemize}

\section{Conclusions}

\noun{Ratracer} shows consistent performance improvement over \noun{Kira}+\noun{FireFly}
(sometimes by an order of magnitude), and sporadic improvement over
\noun{Kira}+\noun{Fermat}. We believe the 5x improvement demonstrated
for challenging reductions such as massive five-point two-loop topologies
is already enough for \noun{Ratracer} to be a valuable option for
problems at the boundary of what is practical.

In the future we envision several ways to improve the performance
of \noun{Ratracer} further. First, seeing that with the fast probe
times that \noun{Ratracer} often provides the overhead in the \noun{FireFly}
library becomes a limiting factor, we see improvements in \noun{FireFly}
to reduce that overhead as an important potential speedup source.
Second, when solving linear equation systems it is possible to achieve
smaller (and therefore faster) traces by spending the time to devise
better equation elimination orders. Third, optimizing the traces further
to make the memory access pattern more linear could improve the performance
of traces with large memory usage.

\subsection*{Acknowledgments}

The author would like to thank Gudrun Heinrich, Sven~Yannick Klein,
Matthias Kerner, and Lisa Biermann for reading through a draft of
this paper and providing comments and suggestions; Sven~Yannick Klein
and Fabian Lange for discussions related to \noun{FireFly} usage
and inner workings. This work was partially funded by the Deutsche
Forschungsgemeinschaft (DFG, German Research Foundation) under grant
396021762---TRR~257.

\printbibliography[heading=bibintoc]

\appendix

\section{Benchmark tables\label{app:benchmark-tables}}

\subsection{Two-loop massive hexatriangle\label{subsec:hexatriangle}}

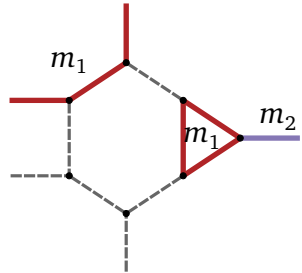
\begin{figure}[H]
\begin{centering}
\raisebox{0.5ex}{\scalebox{1.00}{\begin{tikzpicture}
	\begin{pgfonlayer}{nodelayer}
		\node [style=dot] (0) at (-0.75, -0.5) {};
		\node [style=dot] (1) at (0, -1) {};
		\node [style=dot] (2) at (0.75, -0.5) {};
		\node [style=dot] (3) at (0.75, 0.5) {};
		\node [style=dot] (4) at (0, 1) {};
		\node [style=dot] (5) at (-0.75, 0.5) {};
		\node [style=dot] (6) at (1.5, 0) {};
		\node [style=none] (7) at (-1.5, -0.5) {};
		\node [style=none] (8) at (0, -1.75) {};
		\node [style=none] (9) at (2.25, 0) {};
		\node [style=none] (10) at (-1.5, 0.5) {};
		\node [style=none] (11) at (0, 1.75) {};
		\node [style=none] (12) at (-0.75, 1) {$m_1$};
		\node [style=none] (13) at (1, 0) {$m_1$};
		\node [style=none] (14) at (2, 0.25) {$m_2$};
	\end{pgfonlayer}
	\begin{pgfonlayer}{edgelayer}
		\draw [style=edge] (8.center) to (1);
		\draw [style=edge] (7.center) to (0);
		\draw [style=edge] (5) to (0);
		\draw [style=edge] (0) to (1);
		\draw [style=edge] (1) to (2);
		\draw [style=edge] (3) to (4);
		\draw [style=massive2 edge] (9.center) to (6);
		\draw [style=massive1 edge] (10.center) to (5);
		\draw [style=massive1 edge] (5) to (4);
		\draw [style=massive1 edge] (4) to (11.center);
		\draw [style=massive1 edge] (2) to (3);
		\draw [style=massive1 edge] (3) to (6);
		\draw [style=massive1 edge] (6) to (2);
	\end{pgfonlayer}
\end{tikzpicture}}}
\par\end{centering}
\caption{\label{fig:tth2l_b25}Two-loop hexatriangle with one mass, 2 massless
and 3 massive legs, 8 propagators, and the total of 6 scaleless ratios.}
\end{figure}

In this task integrals from a massive two-loop hexatriangle family
of~\ref{fig:tth2l_b25} are reduced to its 168 master integrals.
Reduction is requested for all 9719 integrals with at most one numerator
and one dot. The solvers are asked to use a fixed master integral
basis which is specifically chosen to be $d$-factorizing. Because
the reduction is fairly complicated, the values of all kinematic and
mass parameters are set to numerical values on a particular line in
the parameter space, leaving only the dimensional regulator~$d$
and the line parameter~$\lambda$ for the reconstruction (i.e. 2~parameters
instead of~6). This task simulates the reduction needed to obtain
differential equations for the master integrals along a particular
line. The family itself appears in $q\bar{q}\to t\bar{t}H$ production
at two loops.

Note that we have imposed a 5-hour time limit on each calculation;
the measurement for \noun{Fire6} was aborted because of this limit.
\begin{center}
\begin{tabular}{lrrrrrrr}
\toprule 
 & Total & \noun{FireFly} & Probe & Probe & \noun{FireFly} & Memory & Disk\tabularnewline
 & time & time & time & count & eff. &  & \tabularnewline
\midrule
\midrule 
\noun{Kira}+\noun{FireFly} & 13407 s & 13367 s &   1.2 s & $7.7\,10^{4}$ & 82.8\% & 4.2 GB &  1.6 GB\tabularnewline
\midrule 
\noun{Ratracer} & 2579 s &  2451 s & 59.0 ms & $7.6\,10^{4}$ & 23.0\% & 3.3 GB &  2.2 GB\tabularnewline
\midrule 
\noun{Ratracer}+Scan &  2759 s &  2659 s & 57.1 ms & $5.3\,10^{4}$ & 14.3\% & 2.4 GB &  2.2 GB\tabularnewline
\midrule 
\noun{Ratracer}+$\mathcal{O}(\varepsilon^{0})$ & 799 s &   697 s & 75.0 ms & $6.1\,10^{3}$ &  8.2\% & 2.0 GB &  2.4 GB\tabularnewline
\midrule 
\noun{Ratracer}+$\mathcal{O}(\varepsilon^{0})$+Scan &  3368 s &  3268 s & 74.5 ms & $1.7\,10^{4}$ &  4.7\% & 2.4 GB &  2.2 GB\tabularnewline
\midrule 
\noun{Ratracer}+$\mathcal{O}(\varepsilon^{1})$ &  1088 s &   970 s & 90.6 ms & $6.1\,10^{3}$ &  7.1\% & 3.1 GB &  2.2 GB\tabularnewline
\midrule 
\noun{Ratracer}+$\mathcal{O}(\varepsilon^{1})$+Scan &  4906 s &  4789 s & 88.9 ms & $1.7\,10^{4}$ &  3.9\% & 3.7 GB &  2.2 GB\tabularnewline
\midrule 
\noun{Ratracer}+$\mathcal{O}(\varepsilon^{2})$ &  1324 s &  1193 s &  106 ms & $6.1\,10^{3}$ &  6.7\% & 4.0 GB &  2.4 GB\tabularnewline
\midrule 
\noun{Ratracer}+$\mathcal{O}(\varepsilon^{2})$+Scan &  6032 s &  5905 s &  103 ms & $1.7\,10^{4}$ &  3.6\% & 4.7 GB &  2.4 GB\tabularnewline
\midrule 
\noun{Kira} & 3156 s & --- & --- & --- & --- & 4.4 GB &  1.4 GB\tabularnewline
\midrule 
\noun{Fire6}+Hints & crash & --- & --- & --- & --- & --- & ---\tabularnewline
\midrule 
\noun{Fire6} & >5 h & --- & --- & --- & --- & >2 GB & >90 GB\tabularnewline
\bottomrule
\end{tabular}
\par\end{center}

\subsection{Non-planar two-loop box with two masses\label{subsec:xbox2l2m}}

\begin{figure}[H]
\begin{centering}
\raisebox{0.5ex}{\scalebox{1.00}{\begin{tikzpicture}
	\begin{pgfonlayer}{nodelayer}
		\node [style=none] (0) at (-1.5, -0.75) {};
		\node [style=none] (1) at (-1.5, 0.75) {};
		\node [style=dot] (2) at (-0.75, -0.75) {};
		\node [style=dot] (3) at (-0.75, 0.75) {};
		\node [style=dot] (4) at (0, -0.75) {};
		\node [style=dot] (5) at (0, 0.75) {};
		\node [style=dot] (6) at (0.75, -0.75) {};
		\node [style=dot] (7) at (0.75, 0.75) {};
		\node [style=none] (8) at (1.5, -0.75) {};
		\node [style=none] (9) at (1.5, 0.75) {};
		\node [style=none] (10) at (1.25, 0.5) {$m_1$};
		\node [style=none] (11) at (-1, 0.25) {$m_2$};
	\end{pgfonlayer}
	\begin{pgfonlayer}{edgelayer}
		\draw [style=edge] (1.center) to (3);
		\draw [style=edge] (0.center) to (2);
		\draw [style=edge] (2) to (4);
		\draw [style=edge] (2) to (5);
		\draw [style=edge] (4) to (6);
		\draw [style=edge] (6) to (7);
		\draw [style=edge] (6) to (8.center);
		\draw [style=massive1 edge] (3) to (5);
		\draw [style=massive1 edge] (5) to (7);
		\draw [style=massive1 edge] (7) to (9.center);
		\draw [style=massive2 edge] (3) to (4);
	\end{pgfonlayer}
\end{tikzpicture}}}
\par\end{centering}
\caption{\label{fig:xbox2l2m}Non-planar 2-loop box with 2 masses, 4 massless
legs, and 7 propagators, and the total of 3 scaleless ratios.}
\end{figure}
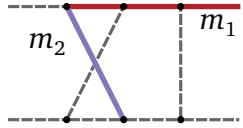

In this task integrals from a two-loop non-planar massive box family
from \ref{fig:xbox2l2m} are reduced to its 76 master integrals. Reduction
is requested for all 3177 integrals with up to two numerators and
no dots. All the solvers are asked to use the same master integrals
as proposed by \noun{Kira} (these have only numerators and no dots).
In~\cite{Klappert:2020nbg} this family is called ``\code{topo5}''.
\begin{center}
\begin{tabular}{lrrrrrrr}
\toprule 
 & Total & \noun{FireFly} & Probe & Probe & \noun{FireFly} & Memory & Disk\tabularnewline
 & time & time & time & count & eff. &  & \tabularnewline
\midrule
\midrule 
\noun{Kira}+\noun{FireFly} & 2482 s & 2470 s &  182 ms & $1.0\,10^{5}$ & 95.9\% & 2.1 GB & 74.8 MB\tabularnewline
\midrule 
\noun{Ratracer} &  713 s &  692 s & 23.4 ms & $1.7\,10^{5}$ & 72.3\% & 1.8 GB &  489 MB\tabularnewline
\midrule 
\noun{Ratracer}+Scan & 412 s &  393 s & 23.3 ms & $1.0\,10^{5}$ & 77.0\% & 1.3 GB &  413 MB\tabularnewline
\midrule 
\noun{Ratracer}+$\mathcal{O}(\varepsilon^{0})$ &  190 s &  163 s & 47.8 ms & $2.4\,10^{4}$ & 89.3\% & 865 MB &  2.7 GB\tabularnewline
\midrule 
\noun{Ratracer}+$\mathcal{O}(\varepsilon^{0})$+Scan &  153 s &  126 s & 47.6 ms & $1.7\,10^{4}$ & 81.9\% & 720 MB &  2.7 GB\tabularnewline
\midrule 
\noun{Ratracer}+$\mathcal{O}(\varepsilon^{1})$ &  382 s &  351 s & 58.5 ms & $4.2\,10^{4}$ & 88.2\% & 1.3 GB &  2.8 GB\tabularnewline
\midrule 
\noun{Ratracer}+$\mathcal{O}(\varepsilon^{1})$+Scan &  299 s &  269 s & 58.0 ms & $3.0\,10^{4}$ & 80.2\% & 1.2 GB &  2.7 GB\tabularnewline
\midrule 
\noun{Ratracer}+$\mathcal{O}(\varepsilon^{2})$ &  710 s &  673 s & 70.1 ms & $6.6\,10^{4}$ & 86.3\% & 2.2 GB &  2.9 GB\tabularnewline
\midrule 
\noun{Ratracer}+$\mathcal{O}(\varepsilon^{2})$+Scan &  532 s &  497 s & 69.7 ms & $4.6\,10^{4}$ & 80.6\% & 1.8 GB &  2.9 GB\tabularnewline
\midrule 
\noun{Kira} & 15199 s & --- & --- & --- & --- & 9.0 GB & 549 MB\tabularnewline
\midrule 
\noun{Fire6}+Hints & >5 h & --- & --- & --- & --- & >4 GB & >28 GB\tabularnewline
\midrule 
\noun{Fire6} & >5 h & --- & --- & --- & --- & >4 GB & >28 GB\tabularnewline
\bottomrule
\end{tabular}
\par\end{center}

\subsection{Two-loop massive pentabubble\label{subsec:pentabubble}}

\begin{figure}[H]
\begin{centering}
\raisebox{0.5ex}{\scalebox{1.00}{\begin{tikzpicture}
	\begin{pgfonlayer}{nodelayer}
		\node [style=none] (0) at (-1.25, 0.5) {};
		\node [style=dot] (1) at (-0.5, 0.5) {};
		\node [style=none] (2) at (1, 1.25) {};
		\node [style=dot] (3) at (0.5, 0.75) {};
		\node [style=dot] (4) at (-0.5, -0.5) {};
		\node [style=none] (5) at (-1.25, -0.5) {};
		\node [style=dot] (6) at (0.5, -0.75) {};
		\node [style=none] (7) at (1, -1.25) {};
		\node [style=dot] (8) at (1, 0) {};
		\node [style=none] (9) at (1.75, 0) {};
		\node [style=none] (10) at (2.25, 0) {$m_2$};
		\node [style=none] (11) at (1.25, -1) {$m_1$};
		\node [style=none] (12) at (1.25, 1) {$m_1$};
	\end{pgfonlayer}
	\begin{pgfonlayer}{edgelayer}
		\draw [style=edge] (1) to (0.center);
		\draw [style=massive1 edge] (3) to (2.center);
		\draw [style=massive2 edge] (8) to (9.center);
		\draw [style=massive1 edge] (6) to (7.center);
		\draw [style=edge] (4) to (5.center);
		\draw [style=massive1 edge, bend right] (1) to (3);
		\draw [style=massive1 edge, bend left=330] (3) to (1);
		\draw [style=massive1 edge] (3) to (8);
		\draw [style=massive1 edge] (8) to (6);
		\draw [style=edge] (1) to (4);
		\draw [style=edge] (4) to (6);
	\end{pgfonlayer}
\end{tikzpicture}}}
\par\end{centering}
\caption{\label{fig:tth2l_b16}Two-loop pentabubble with one mass, 2 massless
and 3 massive legs, 6 propagators, and the total of 6 scaleless ratios.}
\end{figure}
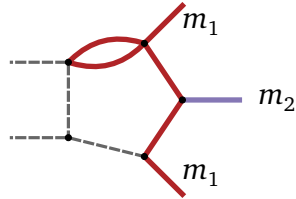

In this task integrals from a massive two-loop hexatriangle family
of \ref{fig:tth2l_b16} are reduced to its 38 master integrals. Reduction
is requested for all 2196 integrals with at most one numerator and
one dot. The solvers are asked to use a fixed master integral basis
which is specifically chosen to be $d$-factorizing; this basis change
is essential for performance in this reduction, as a naive Laporta
basis increases the reduction time by orders of magnitude.
\begin{center}
\begin{tabular}{lrrrrrrr}
\toprule 
 & Total & \noun{FireFly} & Probe & Probe & \noun{FireFly} & Memory & Disk\tabularnewline
 & time & time & time & count & eff. &  & \tabularnewline
\midrule
\midrule 
\noun{Kira}+\noun{FireFly} &   231 s &  224 s & 4.7 ms & $1.5\,10^{5}$ & 40.3\% & 741 MB & 20.1 MB\tabularnewline
\midrule 
\noun{Ratracer} &   210 s &  204 s & 555 us & $1.9\,10^{5}$ &  6.5\% & 1.0 GB & 42.3 MB\tabularnewline
\midrule 
\noun{Ratracer}+Scan &   112 s &  106 s & 531 us & $1.2\,10^{5}$ &  7.5\% & 537 MB & 30.2 MB\tabularnewline
\midrule 
\noun{Ratracer}+$\mathcal{O}(\varepsilon^{0})$ &  20.3 s & 15.0 s & 578 us & $3.5\,10^{4}$ & 16.6\% & 501 MB & 20.1 MB\tabularnewline
\midrule 
\noun{Ratracer}+$\mathcal{O}(\varepsilon^{0})$+Scan &  27.1 s & 22.2 s & 571 us & $2.8\,10^{4}$ &  9.1\% & 374 MB & 18.7 MB\tabularnewline
\midrule 
\noun{Ratracer}+$\mathcal{O}(\varepsilon^{1})$ &  28.1 s & 22.5 s & 686 us & $3.5\,10^{4}$ & 13.2\% & 818 MB & 29.0 MB\tabularnewline
\midrule 
\noun{Ratracer}+$\mathcal{O}(\varepsilon^{1})$+Scan &  41.6 s & 36.1 s & 691 us & $2.9\,10^{4}$ &  6.9\% & 608 MB & 26.8 MB\tabularnewline
\midrule 
\noun{Ratracer}+$\mathcal{O}(\varepsilon^{2})$ &  32.8 s & 26.8 s & 767 us & $3.5\,10^{4}$ & 12.4\% & 971 MB & 33.5 MB\tabularnewline
\midrule 
\noun{Ratracer}+$\mathcal{O}(\varepsilon^{2})$+Scan &  50.6 s & 44.8 s & 773 us & $2.9\,10^{4}$ &  6.2\% & 741 MB & 31.0 MB\tabularnewline
\midrule 
\noun{Kira} &  21.8 s & --- & --- & --- & --- & 1.1 GB & 14.7 MB\tabularnewline
\midrule 
\noun{Fire6}+Hints & crash & --- & --- & --- & --- & --- & ---\tabularnewline
\midrule 
\noun{Fire6} & >5 h & --- & --- & --- & --- & >4 GB & >36 GB\tabularnewline
\bottomrule
\end{tabular}
\par\end{center}

\subsection{Three-loop diamond with two masses\label{subsec:diamond3l}}

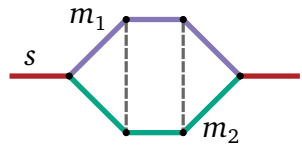
\begin{figure}[H]
\begin{centering}
\raisebox{0.5ex}{\scalebox{1.00}{\begin{tikzpicture}
	\begin{pgfonlayer}{nodelayer}
		\node [style=dot] (0) at (-1, 0) {};
		\node [style=dot] (1) at (-0.25, 0.75) {};
		\node [style=dot] (2) at (-0.25, -0.75) {};
		\node [style=dot] (5) at (0.5, 0.75) {};
		\node [style=dot] (6) at (0.5, -0.75) {};
		\node [style=dot] (7) at (1.25, 0) {};
		\node [style=none] (8) at (-1.75, 0) {};
		\node [style=none] (9) at (2, 0) {};
		\node [style=none] (10) at (-1.5, 0.25) {$s$};
		\node [style=none] (12) at (-0.75, 0.75) {$m_1$};
		\node [style=none] (13) at (1, -0.75) {$m_2$};
	\end{pgfonlayer}
	\begin{pgfonlayer}{edgelayer}
		\draw [style=massive2 edge] (5) to (7);
		\draw [style=massive3 edge] (7) to (6);
		\draw [style=massive3 edge] (2) to (0);
		\draw [style=massive2 edge] (0) to (1);
		\draw [style=massive1 edge] (0) to (8.center);
		\draw [style=massive1 edge] (7) to (9.center);
		\draw [style=edge] (1) to (2);
		\draw [style=edge] (5) to (6);
		\draw [style=massive2 edge] (1) to (5);
		\draw [style=massive3 edge] (2) to (6);
	\end{pgfonlayer}
\end{tikzpicture}}}
\par\end{centering}
\caption{\label{fig:diamond3l}Massive 3-loop diamond with 2 masses, 2 massive
legs, 8 propagators, and the total of 2 scaleless ratios.}
\end{figure}

In this task integrals from a three-loop massive diamond family from
\ref{fig:diamond3l} are reduced to its 40 master integrals. Reduction
is requested for all 24777 integrals with up to two numerators and
one dot.
\begin{center}
\begin{tabular}{lrrrrrrr}
\toprule 
 & Total & \noun{FireFly} & Probe & Probe & \noun{FireFly} & Memory & Disk\tabularnewline
 & time & time & time & count & eff. &  & \tabularnewline
\midrule
\midrule 
\noun{Kira}+\noun{FireFly} &  116 s & 99.7 s & 57.8 ms & $3.3\,10^{3}$ & 23.8\% & 2.8 GB &  116 MB\tabularnewline
\midrule 
\noun{Ratracer} & 69.1 s & 48.2 s &  2.3 ms & $6.7\,10^{3}$ &  3.9\% & 4.1 GB &  184 MB\tabularnewline
\midrule 
\noun{Ratracer}+Scan & 72.6 s & 55.7 s &  2.2 ms & $4.3\,10^{3}$ &  2.2\% & 3.7 GB &  116 MB\tabularnewline
\midrule 
\noun{Ratracer}+$\mathcal{O}(\varepsilon^{0})$ & 29.7 s & 12.6 s &  3.2 ms & $4.1\,10^{2}$ &  1.3\% & 3.7 GB &  103 MB\tabularnewline
\midrule 
\noun{Ratracer}+$\mathcal{O}(\varepsilon^{0})$+Scan & 42.6 s & 25.9 s &  3.3 ms & $6.7\,10^{2}$ &  1.1\% & 3.7 GB & 99.5 MB\tabularnewline
\midrule 
\noun{Ratracer}+$\mathcal{O}(\varepsilon^{1})$ & 41.7 s & 21.7 s &  5.8 ms & $4.1\,10^{2}$ &  1.4\% & 4.9 GB &  159 MB\tabularnewline
\midrule 
\noun{Ratracer}+$\mathcal{O}(\varepsilon^{1})$+Scan & 66.3 s & 47.0 s &  5.1 ms & $6.6\,10^{2}$ &  0.9\% & 3.8 GB &  154 MB\tabularnewline
\midrule 
\noun{Ratracer}+$\mathcal{O}(\varepsilon^{2})$ & 51.1 s & 28.5 s &  6.7 ms & $4.1\,10^{2}$ &  1.2\% & 6.7 GB &  208 MB\tabularnewline
\midrule 
\noun{Ratracer}+$\mathcal{O}(\varepsilon^{2})$+Scan & 84.9 s & 63.2 s &  6.8 ms & $6.8\,10^{2}$ &  0.9\% & 5.0 GB &  202 MB\tabularnewline
\midrule 
\noun{Kira} & 21.3 s & --- & --- & --- & --- & 858 MB &  118 MB\tabularnewline
\midrule 
\noun{Fire6}+Hints &  134 s & --- & --- & --- & --- & 749 MB & 11.4 GB\tabularnewline
\midrule 
\noun{Fire6} &  112 s & --- & --- & --- & --- & 805 MB & 11.4 GB\tabularnewline
\bottomrule
\end{tabular}
\par\end{center}

\subsection{Massive two-loop box\label{subsec:box2l}}

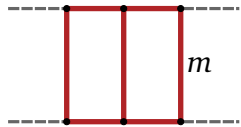
\begin{figure}[H]
\begin{centering}
\raisebox{0.5ex}{\scalebox{1.00}{\begin{tikzpicture}
	\begin{pgfonlayer}{nodelayer}
		\node [style=none] (0) at (-1.5, 0.75) {};
		\node [style=none] (1) at (-1.5, -0.75) {};
		\node [style=dot] (2) at (-0.75, 0.75) {};
		\node [style=dot] (3) at (-0.75, -0.75) {};
		\node [style=dot] (4) at (0, 0.75) {};
		\node [style=dot] (5) at (0, -0.75) {};
		\node [style=dot] (6) at (0.75, 0.75) {};
		\node [style=dot] (7) at (0.75, -0.75) {};
		\node [style=none] (8) at (1.5, 0.75) {};
		\node [style=none] (9) at (1.5, -0.75) {};
		\node [style=none] (10) at (1, 0) {$m$};
	\end{pgfonlayer}
	\begin{pgfonlayer}{edgelayer}
		\draw [style=edge] (0.center) to (2);
		\draw [style=edge] (1.center) to (3);
		\draw [style=edge] (6) to (8.center);
		\draw [style=edge] (7) to (9.center);
		\draw [style=massive1 edge] (2) to (4);
		\draw [style=massive1 edge] (4) to (6);
		\draw [style=massive1 edge] (6) to (7);
		\draw [style=massive1 edge] (7) to (5);
		\draw [style=massive1 edge] (5) to (3);
		\draw [style=massive1 edge] (3) to (2);
		\draw [style=massive1 edge] (4) to (5);
	\end{pgfonlayer}
\end{tikzpicture}}}
\par\end{centering}
\caption{\label{fig:box2l}A two-loop box with one mass, 4 massless legs, 7
propagators, and the total of 2 scaleless ratios.}
\end{figure}

In this task, integrals from a family of the massive two-loop box
from \ref{fig:box2l} are reduced to master integrals. Reduction is
requested for all 35097 integrals with up to two numerators and two
dots.
\begin{center}
\begin{tabular}{lrrrrrrr}
\toprule 
 & Total & \noun{FireFly} & Probe & Probe & \noun{FireFly} & Memory & Disk\tabularnewline
 & time & time & time & count & eff. &  & \tabularnewline
\midrule
\midrule 
\noun{Kira}+\noun{FireFly} &  3933 s & 3899 s &  225 ms & $6.3\,10^{4}$ & 45.8\% & 28.8 GB &  3.1 GB\tabularnewline
\midrule 
\noun{Ratracer} &   579 s &  518 s & 23.9 ms & $2.3\,10^{4}$ & 11.6\% & 14.2 GB &  1.0 GB\tabularnewline
\midrule 
\noun{Ratracer}+Scan &   435 s &  383 s & 23.4 ms & $1.3\,10^{4}$ & 10.2\% & 12.2 GB &  833 MB\tabularnewline
\midrule 
\noun{Ratracer}+$\mathcal{O}(\varepsilon^{0})$ &   101 s & 61.5 s & 23.3 ms & $2.5\,10^{3}$ & 11.8\% & 12.3 GB &  1.3 GB\tabularnewline
\midrule 
\noun{Ratracer}+$\mathcal{O}(\varepsilon^{0})$+Scan &   134 s & 95.6 s & 22.6 ms & $2.3\,10^{3}$ &  6.9\% & 12.4 GB &  1.3 GB\tabularnewline
\midrule 
\noun{Ratracer}+$\mathcal{O}(\varepsilon^{1})$ &   189 s &  136 s & 38.7 ms & $2.8\,10^{3}$ &  9.9\% & 12.6 GB &  1.4 GB\tabularnewline
\midrule 
\noun{Ratracer}+$\mathcal{O}(\varepsilon^{1})$+Scan &   256 s &  205 s & 37.7 ms & $2.6\,10^{3}$ &  5.9\% & 12.1 GB &  1.4 GB\tabularnewline
\midrule 
\noun{Ratracer}+$\mathcal{O}(\varepsilon^{2})$ &   272 s &  205 s & 50.9 ms & $2.8\,10^{3}$ &  8.6\% & 18.1 GB &  1.5 GB\tabularnewline
\midrule 
\noun{Ratracer}+$\mathcal{O}(\varepsilon^{2})$+Scan &   366 s &  304 s & 49.4 ms & $2.6\,10^{3}$ &  5.2\% & 15.5 GB &  1.5 GB\tabularnewline
\midrule 
\noun{Kira} &  1126 s & --- & --- & --- & --- &  5.8 GB &  1.9 GB\tabularnewline
\midrule 
\noun{Fire6}+Hints & >5 h & --- & --- & --- & --- & >3 GB & >80 GB\tabularnewline
\midrule 
\noun{Fire6} &  3848 s & --- & --- & --- & --- &  3.9 GB & 21.1 GB\tabularnewline
\bottomrule
\end{tabular}
\par\end{center}

\end{document}